\documentclass[twocolumn,superscriptaddress,preprintnumbers]{revtex4}
\usepackage{amssymb,epsf,epsfig}

\newcommand{\bary}{\begin{array}}
\newcommand{\eary}{\end{array}}
\newcommand{\be}{\begin{equation}}
\newcommand{\ee}{\end{equation}}
\newcommand{\bea}{\begin{eqnarray}}
\newcommand{\eea}{\end{eqnarray}}
\newcommand{\nn}{\nonumber}
\newcommand{\lsim}{
\mathrel{\hbox{\rlap{\hbox{\lower4pt\hbox{$\sim$}}}\hbox{$<$}}}}
\newcommand{\gsim}{
\mathrel{\hbox{\rlap{\hbox{\lower4pt\hbox{$\sim$}}}\hbox{$>$}}}}

\def\lbar{\overline}

\begin{document}

\preprint{
  \vbox{
    \hbox{MADPH-04-1381}
    \hbox{UPR-1077T}
    \hbox{hep-ph/0406126}
  }}

\title{Solution to the $B \to \pi K$ Puzzle in a Flavor-Changing $Z'$ Model}

\author{Vernon Barger}
\affiliation{Department of Physics, University of Wisconsin, Madison, WI 53706}
\author{Cheng-Wei Chiang}
\affiliation{Department of Physics, University of Wisconsin, Madison, WI 53706}
\author{Paul Langacker}
\affiliation{Department of Physics and Astronomy,
University of Pennsylvania, Philadelphia, PA 19104}
\author{Hye-Sung Lee}
\affiliation{Department of Physics, University of Wisconsin, Madison, WI 53706}

\date{\today}

\begin{abstract}
  \vspace{0.2cm}\noindent Recent experiments suggest that certain $B \to \pi K$
  branching ratios are inconsistent with the standard model expectations.  We
  show that a flavor-changing $Z'$ provides a solution to the problem.
  Electroweak penguin amplitudes are enhanced by the $Z'$ boson for select
  parameters.  We discuss implications for the $Z'$ mass and its couplings to
  the standard model fermions.  We also show that the solution is consistent
  with constraints from the $CP$ asymmetries of the $B \to \phi K_S$ decay.
\end{abstract}

\maketitle

\section{Introduction}

Recent experimental data show that some hadronic $B$ decays to two pseudoscalar
mesons deviate from the standard model (SM) expectations.  In $B \to \pi\pi$
decays, for example, the $\pi^0\pi^0$ mode is found to have a larger branching
ratio than expected \cite{Aubert:2003hf,Chao:2003ue} and the $\pi^+ \pi^-$ mode
has a large direct $CP$ asymmetry \cite{BaBarPlot0054,Abe:2004us}.  It was
subsequently shown that agreement between theory and experiment on the above
observables can be reconciled with a large color-suppressed tree amplitude
($C$), that has a non-trivial strong phase relative to the color-allowed tree
amplitude ($T$), along with a gluonic penguin amplitude ($P$) that also has a
sizeable strong phase relative to $T$ \cite{Buras:0312259,Chiang:2004nm}.  Such
large final-state rescattering effects cannot be easily accounted for in
perturbative approaches \cite{BN,Beneke:2003zv,Keum:2003qi}.  A recent
soft-collinear effective theory analysis \cite{Bauer:2004tj}, however, finds
one solution that can explain the $\pi \pi$ data without a large strong phase
between $T$ and $C$.

Using the flavor $SU(3)$ symmetry, the contributing amplitudes extracted from
the strangeness conserving ($\Delta S = 0$) $\pi \pi$ decays can be related to
those in the strangeness changing ($|\Delta S| = 1$) $\pi K$ decays, allowing
predictions to be made for the $\pi K$ modes within the $SU(3)$ framework.
Most of the $CP$ asymmetries are consistent with the SM predictions, except for
the direct $CP$ asymmetry of the $\pi^0 K^0$ mode, which has recently been
measured but still has a sizeable uncertainty \cite{Aubert:2004xf}.  A
perplexing pattern, however, is observed in the measured $\pi K$ branching
ratios
\cite{Buras:2000gc,Yoshikawa:2003hb,Gronau:2003kj,Beneke:2003zv,Buras:0312259}.
The $\pi K$ puzzle can be stated in terms of the ratios \cite{Buras:epjc.11.93}
\bea
\label{eqn:Rc}
R_{\rm c} &\equiv& 2\left[\frac{\mbox{BR}(B^+\to\pi^0K^+)+
\mbox{BR}(B^-\to\pi^0K^-)}{\mbox{BR}(B^+\to\pi^+K^0)+
\mbox{BR}(B^-\to\pi^-\overline{K^0})}\right] ~~~~~ \nn \\
&=& 1.15 \pm 0.12 ~,
\eea
\bea
\label{eqn:Rn}
R_{\rm n} &\equiv& \frac{1}{2}\left[\frac{\mbox{BR}(B^0\to\pi^-K^+)+
\mbox{BR}(\overline{B^0}\to\pi^+K^-)}{\mbox{BR}(B^0\to\pi^0K^0)+
\mbox{BR}(\overline{B^0}\to\pi^0\overline{K^0})}\right] ~~~~ \nn \\
&=& 0.78 \pm 0.10 ~,
\eea
where the experimental values and those given below for the $CP$ asymmetries
are taken from Ref.~\cite{Chiang:2004nm}.  In the SM, these two ratios should
be approximately equal, whereas values in Eq. (\ref{eqn:Rc}) and (\ref{eqn:Rn})
indicate a $2.4 \sigma$ difference.  This deviation may be due to an
underestimation of the $\pi^0$ detection efficiency \cite{Gronau:2003kj}.
However, if the deviation is real, a sizeable new physics amplitude with a
distinct weak phase is required for the $\pi^0 K^0$ and $\pi^0 K^{\pm}$ decays,
which have a significant dependence on the color-allowed electroweak (EW)
penguins \cite{Yoshikawa:2003hb,Buras:0312259}.  A related ratio that is not
sensitive to the color-allowed EW penguins is \cite{Buras:epjc.11.93}
\bea
\label{eqn:R}
R &\equiv &\left[\frac{\mbox{BR}(B^0\to\pi^-K^+)+
\mbox{BR}(\overline{B^0}\to\pi^+K^-)}{\mbox{BR}(B^+\to\pi^+K^0)+
\mbox{BR}(B^-\to\pi^-\overline{K^0})}\right]\frac{\tau_{B^+}}{\tau_{B^0_d}} \nn \\
&=& 0.91 \pm 0.07 ~.
\eea
where $\tau_{B^+} / \tau_{B^0_d} = 1.086 \pm 0.017$ was used.  The experimental
value of $R$ is consistent with the SM prediction.

In a recent paper \cite{Barger:plb.580.186}, we showed that in a model with an
extra $U(1)'$ gauge boson it is possible to explain the current discrepancy
between the measured mixing $CP$ asymmetry, $S_{\phi K_S}$, and its SM
prediction based upon the $S_{J/\psi K_S}$ measurements if the $Z'$ has family
non-universal couplings with the SM fermions.  A large flavor-changing $Z'$
coupling between the bottom and strange quarks is experimentally allowed.  Its
implications in the $B_s$ system have been studied in
Ref.~\cite{Barger:2004qc}.  Such models can naturally arise in certain string
constructions \cite{Chaudhuri:npb.456.89}, $E_6$-motivated models
\cite{Nardi:prd.48.1240} and dynamical symmetry breaking models
\cite{Buchalla:prd.53.5185}.

A common feature of the above-mentioned $\pi K$ modes and the $\phi K_S$ modes
is that they are all dominated by $b \to s$ penguin-loop processes.  In this
letter, we demonstrate that the $Z'$ model can also provide the indicated new
physics amplitude contributing to the EW penguin amplitudes in the $\pi K$
decays \cite{Grossman:1999av}, while at the same time explaining the $\phi K_S$
$CP$ asymmetry.

In Section \ref{section:model}, we specify our flavor-changing $Z'$ model, the
relevant parameters involved in the analysis, and how the effective Hamiltonian
responsible for hadronic $B$ decays is modified.  We parametrize in Section
\ref{section:parametrization} the EW penguin amplitudes where the new physics
contributions would be manifest.  In Section \ref{section:BKpi}, the master
formulae used in the analysis of $\pi K$ decays are given along with the
related hadronic parameters.  Section \ref{section:solution} gives our
numerical solutions to the $B \to \pi K$ puzzle.  In Section
\ref{section:BphiK}, we compare with the constraints from current $B \to \phi
K_S$ data.  We find that viable solutions to both problems exist.

\section{Flavor Changing $Z'$ Model}
\label{section:model}

The basic formalism of the family non-universal $Z'$ model with flavor-changing
neutral currents (FCNCs) can be found in Ref.~\cite{Langacker:prd.62.013006}.
In this letter, we adopt the formalism and assumptions that we have introduced
in Ref.~\cite{Barger:plb.580.186} for explaining the observed anomalous
indirect $CP$ asymmetry in $B \to \phi K_S$.  For simplicity, we consider the
example in which flavor-changing effects are sizeable only in the left-handed
couplings, leaving the right-handed couplings flavor-diagonal.

We write the $Z'$ part of the neutral-current Lagrangian in the gauge basis as
\be
{\cal L}^{Z'} = - g_2 J'_{\mu} Z'^{\mu} ~,
\ee
where $g_2$ is the gauge coupling constant of the $U(1)'$ group at the $M_W$
scale.  We neglect its renormalization group (RG) running between the $M_W$ and
$M_{Z'}$ scales in view of uncertainties in the parameters.  The $Z'$ chiral
current is
\be
J'_{\mu}
= \sum_{i,j} {\lbar \psi_i^I} \gamma_{\mu}
  \left[ (\epsilon_{\psi_L})_{ij} P_L + (\epsilon_{\psi_R})_{ij} P_R \right]
  \psi^I_j ~,
\ee
where the sum extends over all flavors of the SM fermion fields, the chirality
projection operators are $P_{L,R} \equiv (1 \mp \gamma_5) / 2$, the superscript
$I$ refers to the gauge interaction eigenstates, and $\epsilon_{\psi_L}$
($\epsilon_{\psi_R}$) denotes the left-handed (right-handed) chiral coupling
matrix.  $\epsilon_{\psi_L}$ and $\epsilon_{\psi_R}$ are hermitian by the
requirement of a real Lagrangian.  The mass eigenstates of the chiral fields
are $\psi_{L,R} = V_{\psi_{L,R}} \psi_{L,R}^I$ and the usual CKM matrix is
given by $V_{\rm CKM} = V_{u_L} V_{d_L}^{\dagger}$.  The chiral $Z'$ coupling
matrices in the physical basis of up-type and down-type quarks are,
respectively,
\be
B^X_u
\equiv V_{u_X} \epsilon_{u_X} V_{u_X}^{\dagger} ~, ~~
B^X_d
\equiv V_{d_X} \epsilon_{d_X} V_{d_X}^{\dagger} ~, ~~
(X = L,R)
\ee
where $B^X_{u(d)}$ are hermitian.  As long as the $\epsilon$ matrices are not
proportional to the identity, the $B$ matrices will have non-zero off-diagonal
elements that induce FCNC interactions at tree level.  Our assumption of
flavor-diagonal right-handed couplings demands $B^R_{u(d)} \propto I$.
However, the flavor-changing left-handed couplings will give new contributions
to the SM operators.  The effective Hamiltonian of the ${\bar b} \to {\bar s} q
{\bar q}$ transitions mediated by the $Z'$ is
\bea
{\cal H}_{\rm eff}^{Z'}
&=&\frac{2 G_F}{\sqrt{2}} \left(\frac{g_2 M_Z}{g_1 M_{Z'}}\right)^2 B^{L*}_{sb}
   ({\bar b}s)_{V-A} \sum_q \left( B^L_{qq} ({\bar q}q)_{V-A} \right. \nn \\
&& \left. + B^R_{qq} ({\bar q}q)_{V+A} \right) + \mbox{h.c.} ~,
\label{eqn:Heff1}
\eea
where $g_1 \equiv e / (\sin\theta_W \cos\theta_W)$, and $B^{L*}_{ij}$ and
$B^{R*}_{ij}$ refer to the complex conjugates of left- and right-handed
effective $Z'$ couplings to the quarks $i$ and $j$ at the weak scale,
respectively.  Here we have suppressed the color indices as they match within
the parentheses.  A complete list of operators and their associated
leading-order (LO) Wilson coefficients \cite{Buchalla:1995vs} evaluated at the
$m_b$ scale can be found in Table \ref{table:wilson}.  For various input
parameters, we take $\alpha_s(M_Z) = 0.118$, $\alpha_{\rm EM} = 1/128$,
$\sin^2\theta_W = 0.23$, $M_W = 80.42$ GeV along with the running quark masses
$m_t = 170$ GeV, $m_b = 4.4$ GeV.

\begin{figure}
\includegraphics{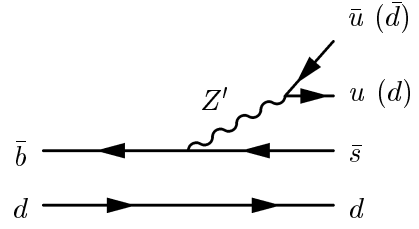}
\caption{A $Z'$-mediated FCNC Feynman diagram contributing to the $B^0 \to
  \pi^0 K^0$ decay.  Its tree-level amplitude with a flavor-changing
  ($Z'$-$b$-$s$) coupling can be sizeable, despite the suppression from the
  large $Z'$ mass.
\label{fig:diagram}}
\end{figure}

The above operators of the forms $({\bar b}s)_{V-A} ({\bar q}q)_{V-A}$ and
$({\bar b}s)_{V-A} ({\bar q}q)_{V+A}$ already exist in the SM, and we can
represent the $Z'$ effect as a modification to the Wilson coefficients of the
corresponding operators.  Thus,
\bea
\lefteqn{{\cal H}_{\rm eff}^{Z'}} \nn \\
&=& - \frac{G_F}{\sqrt{2}} V_{tb}^* V_{ts} \sum_q \left( \Delta c_3 O_3^{(q)} +
  \Delta c_5 O_5^{(q)}  \right. \nn \\
&&\left. + \Delta c_7 O_7^{(q)} + \Delta c_9 O_9^{(q)} \right) + \mbox{h.c.}
\nn \\ 
&=& - \frac{G_F}{\sqrt{2}} V_{tb}^* V_{ts} ({\bar b}s)_{V-A} 
\sum_q \left( ( \Delta c_3 + \Delta c_9 \frac{3}{2} e_q ) ({\bar q}q)_{V-A}
\right.
\nn \\
&& \left.
  + ( \Delta c_5 + \Delta c_7 \frac{3}{2} e_q ) ({\bar q}q)_{V+A} \right) +
\mbox{h.c.} \label{eqn:Heff2}
\eea
The additional contributions to the SM Wilson coefficients at the $M_W$ scale
in terms of $Z'$ parameters from Eq.~(\ref{eqn:Heff1}) and (\ref{eqn:Heff2})
are determined to be
\bea
&&
\Delta c_{3(5)} = - \frac{2}{3 V_{tb}^* V_{ts}} \left(\frac{g_2 M_Z}{g_1
    M_{Z'}}\right)^2 B^{L*}_{sb} \left(B^{L(R)}_{uu} + 2 B^{L(R)}_{dd} \right)
\nn \\
\\
&&
\Delta c_{9(7)} = -\frac{4}{3 V_{tb}^* V_{ts}} \left(\frac{g_2 M_Z}{g_1
    M_{Z'}}\right)^2 B^{L*}_{sb} \left(B^{L(R)}_{uu} - B^{L(R)}_{dd} \right).
\nn \\
\eea
While in general we can have a $Z'$ contribution to the QCD penguins $\Delta
c_{3(5)}$ as well as the EW penguins $\Delta c_{9(7)}$, in view of the results
found by Buras et al. \cite{Buras:0312259} we assume $B^{L(R)}_{uu} \simeq -2
B^{L(R)}_{dd}$ so that new physics is primarily manifest in the EW penguins.
(The same assumption has been used in Ref.~\cite{Barger:plb.580.186}.)  This
can be realized by $Z$-$Z'$ mixing with a small mixing angle $\theta$ at ${\cal
  O}(10^{-3})$, as constrained by current data.  As a result, we have the
$q{\bar q}$ current coupling in part to the $Z'$ with a flavor-universal
coupling and in part to the $Z$ with the SM couplings.  In this case, the
relation $B^{L}_{uu} \simeq -2 B^{L}_{dd}$ can be satisfied when
$\epsilon_{\psi_L} \simeq (\theta g_1 M_{Z'}^2)/(6 g_2 M_Z^2)$ for the quarks
of the first two generations.  The requirement for the corresponding relation
of the right-handed quarks is more relaxed because the $Z'$ couplings to quarks
can be flavor-dependent.

The resulting $Z'$ contributions to the Wilson coefficients at the weak scale
are
\bea
&&\Delta c_{3(5)} \simeq 0 ~, \\
&&\Delta c_{9(7)} = 4 \frac{|V_{tb}^* V_{ts}|}{V_{tb}^* V_{ts}} \xi^{LL(R)}
e^{- i \phi_L} ~,
\label{eqn:c97}
\eea
where
\bea
\xi^{LX} &\equiv& \left(\frac{g_2 M_Z}{g_1 M_{Z'}}\right)^2
\left|\frac{B^{L*}_{sb} B^X_{dd}}{V_{tb}^* V_{ts}}\right| ~~ 
(X=L,R) ~, \label{eqn:xi} \\
\phi_L &\equiv& {\rm Arg}[B^L_{sb}] ~.
\eea
The diagonal elements of the effective coupling matrix $B^X_{ij}$ are real due
to the hermiticity of the effective Hamiltonian, but the off-diagonal elements,
such as $B^L_{sb}$, generally contain new weak phases.

Thus, there are 3 independent real parameters $\{\xi^{LL}$, $\xi^{LR}$,
$\phi_L\}$ in the model considered here.  (Note that a possible negative sign
of $B^X_{dd}$ can be accounted for by shifting $\phi_L$ by $\pi$.)  The
relations $B^{L(R)}_{cc} \simeq B^{L(R)}_{uu}$ and $B^{L(R)}_{ss} \simeq
B^{L(R)}_{dd}$ follow from the assumptions of universality for the first two
families, as required by $K$ and $\mu$ decay constraints
\cite{Langacker:prd.62.013006}.  We assume that the $Z'$ couplings to lepton
pairs ($B^X_{\ell\ell}$) are sufficiently small to satisfy experimental
constraints from the related leptonic decays (i.e., decays to $\ell \bar \ell$
instead of $q \bar q$) \cite{Babu:1996vt}.

The resulting effective Hamiltonian at the $M_W$ scale is then
\bea
{\cal H}_{\rm eff}^{Z'}
&=& - \frac{4 G_F}{\sqrt{2}}
    \left(\frac{g_2 M_Z}{g_1 M_{Z'}}\right)^2 B^{L*}_{sb}
    \sum_q \left( B^L_{dd} O_9^{(q)} + B^R_{dd} O_{7}^{(q)} \right) \nn \\
&&+ \mbox{h.c.}
\eea
Since heavy degrees of freedom (including the $Z'$) in the theory are
considered to have been integrated out at the scale of $M_W$, the RG evolution
of the Wilson coefficients down to low energies after including the new
contributions from $Z'$ is exactly the same as in the SM.

\section{Parametrization of EW Penguin Contribution}
\label{section:parametrization}

\begin{table}[ht]
\caption{The Wilson coefficients for the SM and additions from the $Z'$ at the
  $m_b$ scale.  The notation is $X\equiv \xi^{LL} e^{i \phi_L}$, $Y\equiv 
  \xi^{LR} e^{i \phi_L}$. The numerical values here are rounded-off from the
  values used in the analysis.
\label{table:wilson}}
\begin{ruledtabular}
\begin{tabular}{lrl}
$~~~~~~~~~~~$Operator & $c_i^{SM}(m_b)$ & $\Delta
c_i(m_b)$ \\ 
\hline
$O^{(q)}_1 = (\bar b_\alpha q_\alpha)_{V-A} (\bar q_\beta s_\beta)_{V-A}$
& $ 1.138~~$ & $~~0.0$ \\
$O^{(q)}_2 = (\bar b_\alpha q_\beta)_{V-A} (\bar q_\beta s_\alpha)_{V-A}$
& $-0.296~~$ & $~~0.0$ \\
$O^{(q)}_3 = (\bar b_\alpha s_\alpha)_{V-A} (\bar q_\beta q_\beta)_{V-A}$
& $ 0.014~~$ & $~~0.0$ \\
$O^{(q)}_4 = (\bar b_\alpha s_\beta)_{V-A} (\bar q_\beta q_\alpha)_{V-A}$
& $-0.029~~$ & $- 0.1 X^*$ \\
$O^{(q)}_5 = (\bar b_\alpha s_\alpha)_{V-A} (\bar q_\beta q_\beta)_{V+A}$
& $ 0.008~~$ & $~~0.0$ \\
$O^{(q)}_6 = (\bar b_\alpha s_\beta)_{V-A} (\bar q_\beta q_\alpha)_{V+A}$
& $-0.036~~$ & $- 0.2 X^*$ \\
$O^{(q)}_7 = \frac{3}{2} e_q (\bar b_\alpha s_\alpha)_{V-A} (\bar q_\beta
q_\beta)_{V+A}$     & $ 0.000~~$ & $- 3.6 Y^*$ \\
$O^{(q)}_8 = \frac{3}{2} e_q (\bar b_\alpha s_\beta)_{V-A} (\bar q_\beta
q_\alpha)_{V+A}$     & $ 0.000~~$ & $- 1.5 Y^*$ \\
$O^{(q)}_9 = \frac{3}{2} e_q (\bar b_\alpha s_\alpha)_{V-A} (\bar q_\beta
q_\beta)_{V-A}$     & $-0.010~~$ & $- 4.5 X^*$ \\
$O^{(q)}_{10} = \frac{3}{2} e_q (\bar b_\alpha s_\beta)_{V-A} (\bar q_\beta
q_\alpha)_{V-A}$  & $ 0.002~~$ & $+ 1.2 X^*$
\end{tabular}
\end{ruledtabular}
\end{table}

The ratio between the EW penguin amplitude ($P_{\rm EW}'$) and the tree
contributions ($T'$ and $C'$) in $|\Delta S| = 1$ decays can be given by the
ratio of the corresponding Wilson coefficients
\cite{Neubert:1998pt,Buras:epjc.11.93,Fleischer:epjc.6.451}:
\bea
\frac{P_{\rm EW}'}{T'+C'} =-\frac{3}{2}\frac{1}{\lambda |V_{ub}/V_{cb}|} 
\left[ \frac{\sum_{k=7}^{10} c_k(m_b) \langle {\lbar O}_k(m_b)
    \rangle}{\sum_{j=1}^{2} c_j(m_b) \langle {\lbar O}_j(m_b) \rangle} \right]
~, \nn \\
\label{eqn:ptc}
\eea
The primes here indicate the amplitudes for $|\Delta S| = 1$ transitions.  The
operator ${\lbar O}_k$ and the hadronic matrix element $\langle {\lbar O}_k
(m_b) \rangle$ are defined as
\bea
{\lbar O}_k &\equiv&
\frac12 \left( O_k^{(u)} - O_k^{(d)} \right) ~, \\
\langle {\lbar O}_k \rangle &\equiv& 
\sqrt{2} \langle K^0 \pi^0 | {\lbar O}_k | B^0 \rangle + \langle K^+ \pi^- |
{\lbar O}_k | B^0 \rangle ~.
\eea
We can then rewrite Eq. (\ref{eqn:ptc}) as
\bea
\frac{P_{\rm EW}'}{T'+C'} &=& -\frac{3}{2}\frac{1}{\lambda |V_{ub}/V_{cb}|}
\left[\frac{c_9(m_b)\tilde\xi + c_{10}(m_b)}{c_1(m_b) +
    c_2(m_b)\tilde\xi} \right. \nn \\ 
&& \qquad
\left. + \frac{c_7(m_b)\tilde\eta + c_8(m_b)\tilde\eta'}{c_1(m_b) +
    c_2(m_b)\tilde\xi} \right ] ~,
\eea
with
\be
\tilde\xi \equiv
\frac{\langle {\lbar O}_2(m_b) \rangle}
     {\langle {\lbar O}_1(m_b) \rangle}, \quad 
\tilde\eta \equiv
\frac{\langle {\lbar O}_7(m_b) \rangle}
     {\langle {\lbar O}_1(m_b) \rangle}, \quad 
\tilde\eta' \equiv
\frac{\langle {\lbar O}_8(m_b) \rangle}
     {\langle {\lbar O}_1(m_b) \rangle}. \nn
\ee
Here we have used the fact that $O_{9,10}$ and $O_{2,1}$ are Fierz-equivalent,
respectively, to simplify the above relation.  For consistency, we have to use
the LO Wilson coefficients to estimate the ratio (\ref{eqn:ptc}).  The scale
dependence of this ratio is negligible as one varies the scale from $m_b/2$ to
$2m_b$.  In the exact $SU(3)$ flavor symmetry limit, $\tilde\xi = 1$, which is
used in our numerical analysis.  There is no symmetry argument for the values
of $\tilde\eta$ and $\tilde\eta'$.  Assuming factorization, their magnitudes
should be of ${\cal O}(1)$ too.  In view of uncertainties in other parameters
in the model, we will take $\tilde\eta = \tilde\eta' = -1$ in later analysis,
where the minus sign comes from the sign flip of the axial current in the
${\bar q}q$ pairs in the $O_{7,8}$ operators.

One should note that although $c_{7,8}$ play a less important role compared
with $c_{9,10}$ in the SM, they can receive contributions from the $Z'$ such
that we cannot neglect them.  In Ref.~\cite{Buras:0312259}, it was implicitly
assumed that new physics contributes dominantly to the {\small
  $(V-A)\otimes(V-A)$} EW penguins.  As one of their conclusions under this
assumption, the current $S_{\phi K_S}$ data cannot be accommodated if one wants
to explain the $\pi K$ anomaly.  We will show in Section \ref{section:BphiK}
that both problems can be solved simultaneously once the right-handed currents
are also taken into account.

We summarize here our simplifications to a general $Z'$ model: we assume
(i) no right-handed flavor-changing couplings ($B^R_{ij} = 0$ for $i \ne j$),
(ii) no significant RG running effect between $M_{Z'}$ and $M_W$ scales,
(iii) negligible $Z'$ effect on the QCD penguin ($\Delta c_{3,5} = 0$) so that
the new physics is manifestly isospin violating.
With these simplifications, we have 3 parameters left in the model.  This
approach provides a minimal way to introduce the $Z'$ effect in the $B \to \pi
K$ puzzle.  Of course, more general $Z'$ models are possible.

As shown in Table~\ref{table:wilson}, the Wilson coefficients at the $m_b$
scale, $c_i(m_b)$, are all real in the SM.  Possible new weak phases can come
from new physics coded in the quantities $X$ and $Y$ defined by
\be
X \equiv \xi^{LL} e^{i \phi_L}, \qquad Y \equiv \xi^{LR} e^{i \phi_L}.
\ee
Because of the operator mixing from the RG running, the QCD penguins also
receive contributions from the $Z'$ even though they are assumed to be
unaffected at the weak scale.  Their sizes should not be comparable to their SM
counterparts to guarantee our assumption that new physics is manifest only in
the EW penguin sector.  We numerically checked that they change only up to
several percent from their SM values for $|X| \lsim 0.02$ and/or $|Y| \lsim
0.02$, the natural sizes of the TeV-scale $Z'$ model \cite{Barger:plb.580.186}.
The current-current terms, however, are virtually unaffected.

Following Ref.~\cite{Buras:0312259}, we define $q$ and $\varphi$ as the
magnitude and the weak phase of the ratio in Eq.~(\ref{eqn:ptc}) and ignore a
small strong phase.  From Table~\ref{table:wilson}, the Wolfenstein parameter
\cite{Wolfenstein:prl.51.1983} $\lambda = 0.2240$ and $|V_{ub}/V_{cb}| =
0.086$ \cite{Battaglia:2003in}, we obtain
\bea
q e^{i \varphi} &\equiv& \frac{P_{\rm EW}'}{T'+C'} \nn \\
&\approx& 0.75 \left[1 + 410 X^* + 450 Y^* \tilde\eta + 180 Y^* \tilde\eta'
\right] ~. \label{eqn:qphi}
\eea
Because of the sign difference between $c_{7,8}$ and $c_{9,10}$ terms (coming
from the relative signs between $\tilde\eta^{(\prime)}$ and $\tilde\xi$), where
dominant $Y$ and $X$ contributions respectively enter, $q e^{i \varphi}$ has
contributions of opposite signs from $X$ and $Y$.  In the SM $(X = Y = 0)$, $q
= 0.75$ and $\varphi = 0^\circ$.

\section{$B \to \pi K$ decays}
\label{section:BKpi}

In terms of the hadronic parameters defined in Ref.~\cite{Buras:0312259}, the
$B\to\pi K$ amplitudes are
\bea
A(B^+\to\pi^+K^0)&=&-P', \\
\sqrt{2}A(B^+\to\pi^0K^+)&=&P'\left[1-\left(e^{i\gamma}-qe^{i\varphi}\right)
r_{\rm c}e^{i\delta_{\rm c}}\right], \\
A(B^0_d\to\pi^-K^+)&=&P'\left[1-re^{i\delta}e^{i\gamma}\right], \\
\sqrt{2}A(B^0_d\to\pi^0K^0)&=&-P'\left[1+\rho_{\rm n}e^{i\theta_{\rm n}}
e^{i\gamma}-qe^{i\varphi}r_{\rm c}e^{i\delta_{\rm c}}\right] ~, \nn \\
\eea
where color-suppressed EW penguin and annihilation amplitudes have been
neglected.  As one immediately sees, only the $\pi^0K^+$ and $\pi^0K^0$
modes are sensitive to the color-allowed EW penguins (parametrized
by $q$ and $\varphi$).  The numerical values of the hadronic parameters
extracted from the $\pi \pi$ modes are \cite{Buras:0312259}
\bea
r=0.11^{+0.07}_{-0.05} ~, \quad &&
\delta=+(42^{+23}_{-19})^\circ ~, \nn \\
\rho_{\rm n}=0.13^{+0.07}_{-0.05} ~, \quad &&
\theta_{\rm n}=-(29^{+21}_{-26})^\circ ~, \label{eqn:hadronic} \\
r_{\rm c}=0.20^{+0.09}_{-0.07} ~, \quad &&
\delta_{\rm c}=+(2^{+23}_{-18})^\circ ~. \nn
\eea
We will take the following CKM weak phase values \cite{Battaglia:2003in}
\be
\gamma=(65\pm 7)^\circ ~, \qquad \phi_d \equiv 2\beta=(47\pm4)^\circ ~.
\label{eqn:ckmphase}
\ee
To focus on the discussions of $Z'$ effects, we will use exclusively the
central values of all the above parameters in Eqs.~(\ref{eqn:hadronic}) and
(\ref{eqn:ckmphase}) in the following analysis, with the understanding that the
allowed parameter space and the errors on predicted quantities only reflect
those of Eqs.~ (\ref{eqn:Rc}) $\sim$ (\ref{eqn:Rn}) and should be larger once
the above uncertainties are included.

\begin{figure*}
\includegraphics[width=.45\textwidth]{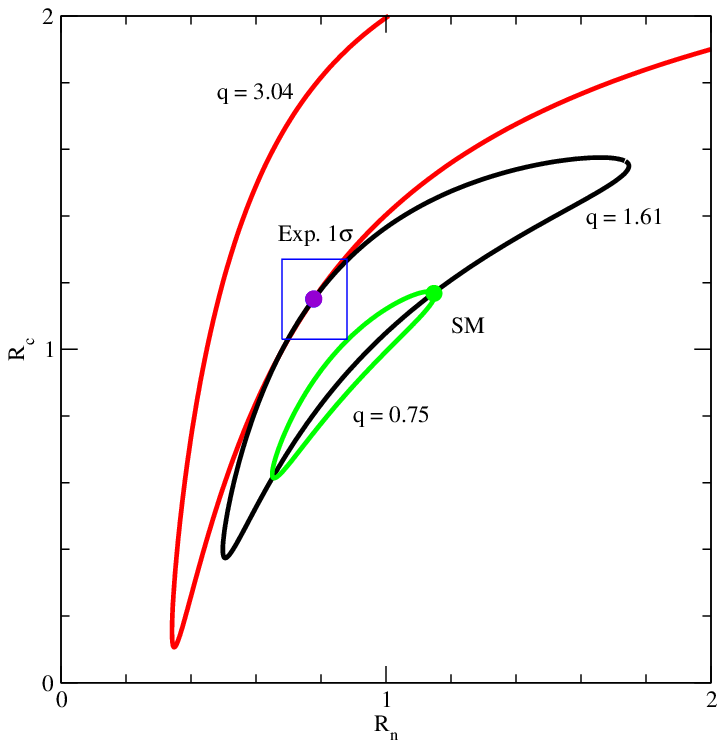} \hspace{0.5cm}
\includegraphics[width=.45\textwidth]{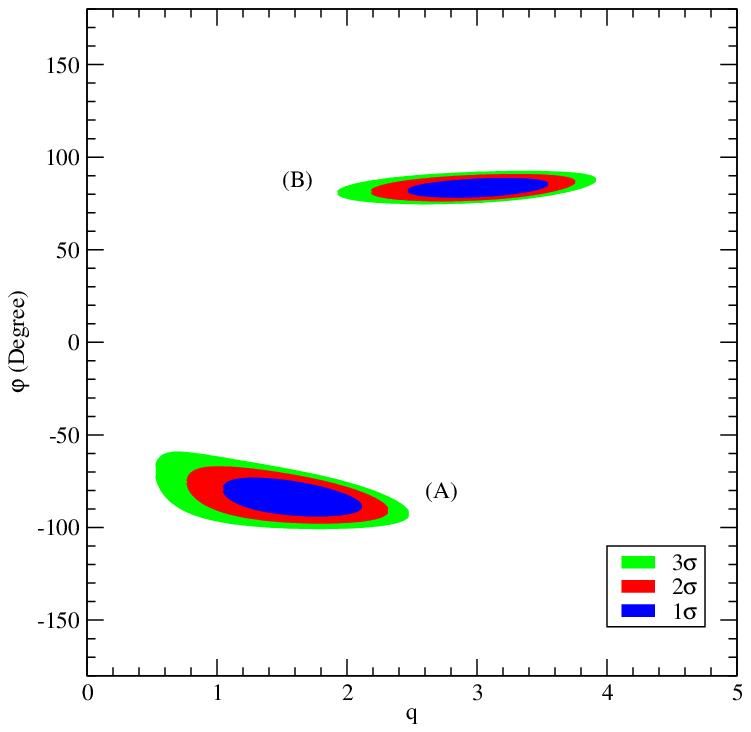} \\
  (a) \hspace{0.45\textwidth} (b)
\caption{(a) Two solutions of $\{q, \varphi\}$ shown on the $R_{\rm n}$-$R_{\rm
    c}$ plane.  Solution (A) $\{q, \varphi\} = \{1.61, -84^\circ\}$ and
  solution (B) $\{q, \varphi\} = \{3.04, 83^\circ\}$ are respectively on the
  black (dark) curve and the red (light) curve traced out by varying $\varphi$
  from $0^\circ$ to $360^\circ$.  The $1\sigma$ experimental bounds and the SM
  prediction for $\{q, \varphi\} = \{0.75, 0^\circ\}$ are also indicated.  (b)
  The contours of solutions (A) and (B) on the $q$-$\varphi$ plane.  Only the
  central values of the hadronic parameters in Eq. (\ref{eqn:hadronic}) are
  used in making these plots.  Note that Solution (A) has been given by
  Ref.~\cite{Buras:0312259}.
\label{fig:twofold}} 
\end{figure*}

The $\pi K$ puzzle can be most easily seen from the $2.4 \sigma$ discrepancy
between the ratios $R_{\rm c}$ and $R_{\rm n}$ in Eqs.~(\ref{eqn:Rc}) and
(\ref{eqn:Rn}) that are predicted to be roughly equal within the SM.  Written
in terms of the above-mentioned hadronic parameters, one obtains
\cite{Buras:0312259}
\bea
R_{\rm c}
&=& 1-2r_{\rm c}\cos\delta_{\rm c}\cos\gamma+r_{\rm c}^2 \nn \\
&& + q r_{\rm c}\left\{ 2\left[ \cos\delta_{\rm c}\cos\varphi
-r_{\rm c}\cos(\gamma-\varphi) \right] + qr_{\rm c} \right\} ~, \nn \\
\label{eqn:Rcwithq} \\
R_{\rm n} &=& \frac{1}{b}\left[1-2r\cos\delta\cos\gamma+r^2\right] ~,
\label{eqn:Rnwithq} 
\eea
with
\bea
b
&\equiv&
1-2qr_{\rm c}\cos\delta_{\rm c}\cos\varphi+q^2r_{\rm c}^2 + \rho_{\rm n}^2
\nn \\
&&
+ 2\rho_{\rm n}\left[\cos\theta_{\rm n}\cos\gamma-qr_{\rm c}\cos(\theta_{\rm
    n}-\delta_{\rm c})\cos(\gamma-\varphi)\right] ~. \nn \\
\eea
Both of these observables are sensitive to the color-allowed EW penguin
amplitudes, as shown by the $q$ and $\varphi$ dependence.  We also set a small
strong phase $\omega = 0$ here and below as found in Ref.~\cite{Buras:0312259}.

Other EW-penguin-sensitive observables are the $CP$ asymmetry of $B^\pm\to\pi^0
K^\pm$ \cite{Buras:0312259}
\bea
A_{CP}(\pi^0 K^\pm)
= - \frac{2}{R_{\rm c}}
  \left[ r_{\rm c}\sin\delta_{\rm c}\sin\gamma
    - qr_{\rm c} \sin\delta_{\rm c}\sin\varphi \right]
\eea
and the time-dependent $CP$ asymmetry of $B \to \pi K_S$
\bea
A_{CP}(t)
= A_{\pi K_S} \cos (\Delta M_{B_d} t) + S_{\pi K_S} \sin (\Delta M_{B_d} t) ~.
\eea
Both the direct and the indirect $CP$ asymmetries have recently been measured
by the BaBar collaboration \cite{Aubert:2004xf}.  In terms of the hadronic
parameters, the asymmetries are given by \cite{Buras:0312259}
\bea
A_{\pi K_S}
&=& -\frac{2}{b}\Bigl[qr_{\rm c}\sin\delta_{\rm c}\sin\varphi -
\rho_{\rm n}\left\{\sin\theta_{\rm n}\sin\gamma \right. \nn \\ 
&& \left. -qr_{\rm c} \sin(\theta_{\rm n}-\delta_{\rm c})
\sin(\gamma-\varphi)\right\}\Bigr] ~, \\
\nn \\
S_{\pi K_S}
&=& \frac{1}{b}\Bigl[\sin\phi_d-2qr_{\rm c}\cos\delta_{\rm c}
\sin(\phi_d+\varphi) \nn \\
&&+q^2r_{\rm c}^2\sin(\phi_d+2\varphi)+2\rho_{\rm n}\left\{\cos\theta_{\rm
    n}\sin(\phi_d+\gamma) \right. \nn \\
&&\left. -qr_{\rm c}\cos(\theta_{\rm n}-\delta_{\rm c})
\sin(\phi_d+\gamma+\varphi)\right\} \nn \\
&&+\rho_{\rm n}^2\sin(\phi_d+2\gamma)\Bigr] ~.
\eea

Using the central values of the parameters given in Eqs.~(\ref{eqn:hadronic})
and (\ref{eqn:ckmphase}) along with $\{q, \varphi\}|_{\mbox{\small SM}} =
\{0.75, 0^\circ\}$, the SM predictions are: $R_{\rm c} = 1.17$, $R_{\rm n} =
1.15$, $A_{CP}(\pi^0 K^\pm) = -0.01$, $A_{\pi K_S} = -0.12$ and $S_{\pi K_S} =
0.86$, which confirm what have been found in Ref.~\cite{Buras:0312259}.

There are other $B \to \pi K$ quantities, including $R$ of Eq.~(\ref{eqn:R}),
that are not sensitive to the EW penguin amplitudes (or the parameters $q$ and
$\varphi$) \cite{Buras:0312259}:
\bea
R &=& 1-2r\cos\delta\cos\gamma+r^2 \\
A_{CP}(\pi^\mp K^\pm) &=& - \frac{2 r \sin\delta \sin\gamma}{1 - 2 r \cos\delta
  \cos\gamma + r^2} \\
A_{CP}(\pi^\pm K^0) &=& 0~.
\eea
Within our assumption of ignoring annihilation amplitudes, $A_{CP} (\pi^\pm
K^0)$ vanishes identically.  Our SM predictions for the other observables are
$R = 0.94$ and $A_{CP}(\pi^\mp K^\pm) = -0.14$, and these values do not change
with the enhancement in the EW penguins by new physics.

\section{Flavor Violating $Z'$ Solution}
\label{section:solution}

\begin{figure*}
  \includegraphics[width=.45\textwidth]{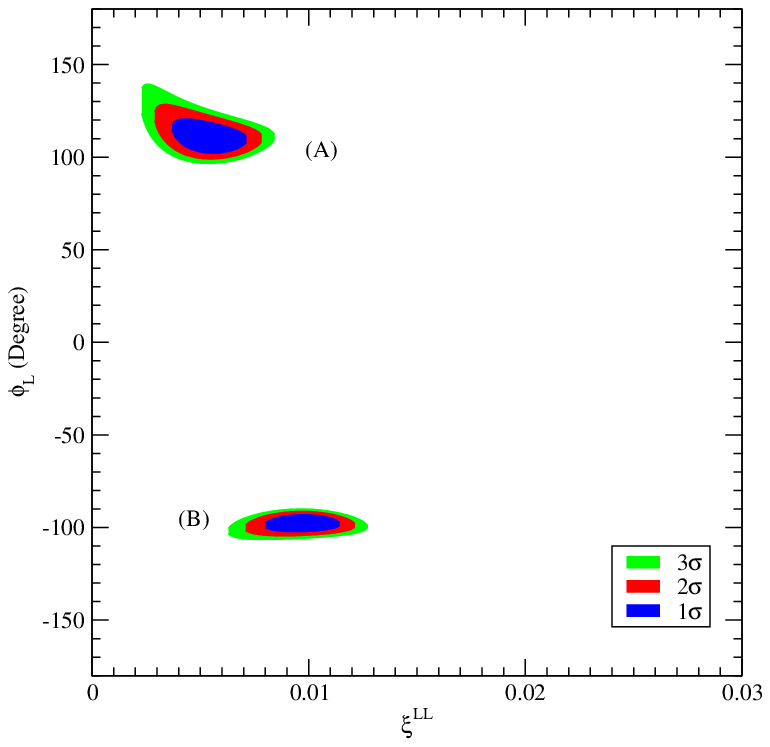} \hspace{0.5cm}
  \includegraphics[width=.45\textwidth]{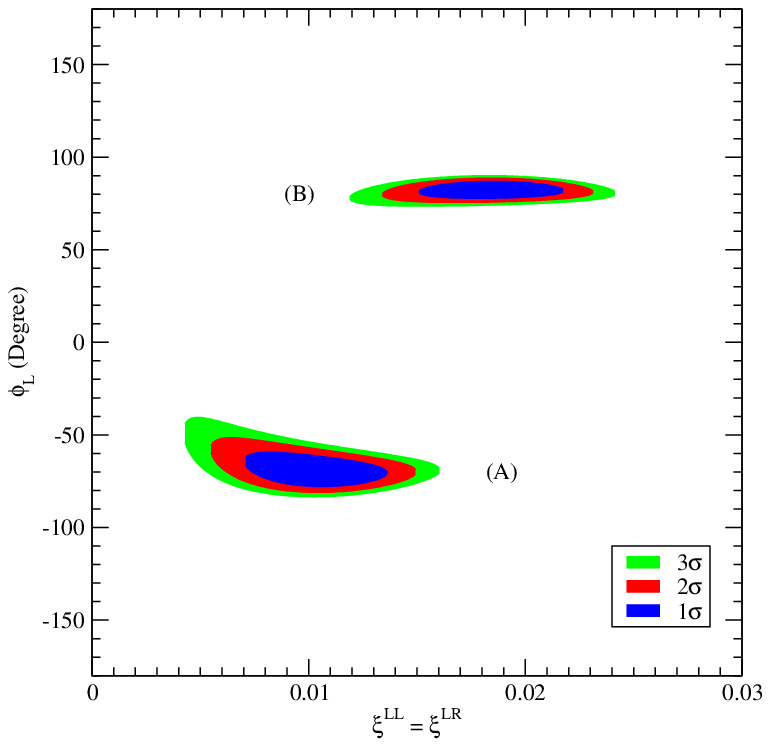} \\
  (a) \hspace{0.45\textwidth} (b)
\caption{Parameter space constrained by the experimental data of $R_{\rm c}$ and
  $R_{\rm n}$, using the central values of the hadronic parameters in
  Eq.~(\ref{eqn:hadronic}) and the CKM phases in Eq.~(\ref{eqn:ckmphase}).  (a)
  Assuming $\xi^{LR} = 0$, solution (A) $\{q, \varphi\} = \{1.61, -84^\circ\}$
  and solution (B) $\{q, \varphi\} = \{3.04, 83^\circ\}$ correspond to
  $\{\xi^{LL},\phi_L\} = \{0.0055,110^\circ\}$ and $\{0.0098,-97^\circ\}$,
  respectively.  (b) Assuming $\xi^{LR} = \xi^{LL}$, solution (A) and solution
  (B) correspond to $\{\xi^{LL},\phi_L\} = \{0.0104,-70^\circ\}$ and
  $\{0.0186,83^\circ\}$, respectively.
\label{fig:contour}}
\end{figure*}

To illustrate that the flavor-changing $Z'$ model can provide the solution to
the $B \to \pi K$ puzzle through the enhanced EW penguin contribution, we take
the central values of the $B \to \pi K$ hadronic parameters in
Eq.~(\ref{eqn:hadronic}) and the SM weak phases in Eq.~(\ref{eqn:ckmphase}).
From the two observables $\{R_{\rm c}$, $R_{\rm n}\}$ given in
Eqs.~(\ref{eqn:Rc}) and (\ref{eqn:Rn}), we find two solutions of the EW penguin
parameters:
\bea
\mbox{(A)}~~~ \{q, \varphi\} &=& \{1.61, -84^\circ\}, \label{eqn:A}\\
\mbox{(B)}~~~ \{q, \varphi\} &=& \{3.04, 83^\circ\}.  \label{eqn:B}
\eea
Solution (A) $\{q, \varphi\} = \{1.61, -84^\circ\}$ is equivalent to the
solution given in Ref.~\cite{Buras:0312259}.  Solution (B) has a larger $q$ and
a $\varphi$ of the opposite sign.  This solution is not ruled out by other
constraints such as rare kaon decays because of different $Z'$ couplings in our
scenario.  Fig.~\ref{fig:twofold} shows the two-fold solutions on the $R_{\rm
  n}$-$R_{\rm c}$ plane and the $q$-$\varphi$ plane.

The solutions for the $Z'$ parameters $\xi^{LL}$, $\xi^{LR}$ and $\phi_L$ in
terms of the general EW penguin parameter $q$ and $\varphi$ can be obtained by
finding the solutions of Eq.~(\ref{eqn:qphi}).  Restricting ourselves to the
case $\xi^{LR} = 0$ (in the limit with only left-handed currents), we find the
solutions
\bea
(\mbox{A}_{L})~~~ \{\xi^{LL}, \phi_L\}
&=& \{0.0055, 110^\circ\} \label{eqn:ALH} ~, \\
(\mbox{B}_{L})~~~ \{\xi^{LL}, \phi_L\}
&=& \{0.0098, -97^\circ\} \label{eqn:BLH} ~.
\eea

As Eq.~(\ref{eqn:xi}) shows, the $\xi^{LL}$ values found above have
implications on the coupling constant and the mass of $Z'$.  In $E_6$ motivated
models, $g_2$ has a value of the same order as $g_1$.  The value of
$|B^{L*}_{sb} B^L_{dd}|$ is unknown, but if we assume it is of the same order
as $|V_{tb} V^*_{ts}| \simeq 0.04$, a size of $\xi^{LL} \sim {\cal O}(0.01)$
would give $M_{Z'} \sim {\cal O}(1 \mbox{ TeV})$, consistent with our model of
TeV-scale $Z'$.

In the special case where $\xi^{LR} = \xi^{LL}$ and $\tilde\eta = \tilde\eta' =
-1$, we find the solutions
\bea
(\mbox{A}_{LR})~~~ \{\xi^{LL} = \xi^{LR}, \phi_L\}
&=& \{0.0104, -70^\circ\} \label{eqn:ARH} ~, \\
(\mbox{B}_{LR})~~~ \{\xi^{LL} = \xi^{LR}, \phi_L\}
&=& \{0.0186,  83^\circ\} \label{eqn:BRH} ~.
\eea
The $180^\circ$ phase change of the solutions $(\mbox{A}_{LR})$ and
$(\mbox{B}_{LR})$ from $(\mbox{A}_{L})$ and $(\mbox{B}_{L})$ can be understood
as a result of the sign change in the dominant new physics term in
Eq.~(\ref{eqn:ptc}).  These solutions are plotted in Fig.~\ref{fig:contour}.
Fig.~\ref{fig:contour} (a) shows the contours of the two fold solutions on the
$\xi^{LL}$-$\phi_L$ plane in the case of $\xi^{LR} = 0$ and
Fig.~\ref{fig:contour} (b) in the case of $\xi^{LR} = \xi^{LL}$.

Using the solutions of $q$ and $\varphi$ found in Eqs. (\ref{eqn:A}) and
(\ref{eqn:B}), we have the following predictions for the $CP$ asymmetries of
various $\pi K$ modes:
\bea
\begin{array}{rcrcr}
A_{CP}(\pi^0 K^\pm)
= &\rm{(A)}&-0.03 \pm 0.01 &\rm{(B)}& 0.03 \pm 0.01 \\
A_{\pi K_S}
= &\rm{(A)}&-0.06 \pm 0.01 &\rm{(B)}&-0.15 \pm 0.01 \\
S_{\pi K_S}
= &\rm{(A)}& 1.00_{-0.02}^{+0.00}~~ & \rm{(B)}&-0.13_{-0.14}^{+0.20}~~ \nn 
\end{array}
\eea
These results are irrespective of whether right-handed currents are included.
The predictions given by Solution (A) are consistent with
Ref.~\cite{Buras:0312259}.  The current experimental values are $A_{CP}(\pi^0
K^\pm) = 0.00 \pm 0.12~ (S=1.79)$, $A_{\pi K_S} = -0.40 \pm 0.29$, and $S_{\pi
  K_S} = 0.48 \pm 0.42$ \cite{Chiang:2004nm}.  Their SM values are already
given in Section \ref{section:BKpi}.  The observables that are not sensitive to
the EW penguin (or parameters $q$, $\varphi$) do not change from their SM
expectations.

\section{Relation to $B \to \phi K_S$ Mode \label{section:BphiK}}

The recent measurement of $S_{\phi K_S} = -0.96\pm0.50^{+0.09}_{-0.11}$ at
Belle showing a $3.5 \sigma$ deviation \cite{Abe:prl.91.261602} from the SM
expectation $0.736 \pm 0.049$ \cite{Browder:2003ii} has aroused great interest
in new physics explanations.  (The BaBar value \cite{Aubert:2004ii}, $0.47 \pm
0.34 ^{+0.08}_{-0.06}$, is consistent with the SM.)  The $Z'$ solution to the
$S_{\phi K_S}$ anomaly has been discussed in our previous paper
\cite{Barger:plb.580.186}.  In this section, we compare our constrained
parameter values extracted in the above sections to those in the $B \to \phi
K_S$ process.  The $\phi K_S$ $CP$ asymmetries can be expressed as
\cite{Buras:0312259}
\bea
A_{\phi K_S}
&=& \frac{2 v_0 \sin\Delta_0 \sin\phi'}
         {1 + 2 v_0 \cos\Delta_0 \cos\phi' + v_0^2} ~, \\
S_{\phi K_S}
&=& \frac{1}{1 + 2v_0\cos\Delta_0\cos\phi' + v_0^2}
    \left[ \,  \sin\phi_d \right. \nn \\
&&  \left. + 2v_0\cos\Delta_0\sin(\phi_d+\phi') 
           + v_0^2\sin(\phi_d+2\phi') \right] ~, \nn \\
\eea
where $v_0 \simeq 0.2$, the weak phase $\phi' = 0$ and the strong phase
$\Delta_0 \simeq \pi$ in the SM.  Assuming factorization, we have
\bea
\lefteqn{v_0 e^{i(\Delta_0 + \phi')}} \nn \\
&\simeq& -\frac{3c_7(m_b)+c_8(m_b)+4(c_9(m_b)+c_{10}(m_b))}
          {8(c_3(m_b)+c_4(m_b))+6c_5(m_b)+2c_6(m_b)}.
\eea
From this equation we can see the $S_{\phi
  K_S}$ receives a roughly symmetric contribution of $c_7$ and $c_9$, and thus
of $X$ and $Y$.

Since the measured $A_{\phi K_S} = 0.05 \pm 0.26$ is consistent with $0$, we
take $\Delta_0 \simeq \pi$ as in the SM.  We can then obtain $v_0$ and $\phi'$
for each solution of the $B \to \pi K$ puzzle from the previous section and
calculate $S_{\phi K_S}$.  For $B \to \pi K$ solutions in the $\xi^{LR} = 0$
case, $S_{\phi K_S} = 0.99$ for solution $(\mbox{A}_{L})$ of
Eq.~(\ref{eqn:ALH}) and $S_{\phi K_S} = -0.69$ for solution $(\mbox{B}_{L})$ of
Eq.~(\ref{eqn:BLH}).  Only the latter, with the larger $q$, satisfies the
world-averaged experimental $1 \sigma$ range, $S_{\phi K_S} = -0.147 \pm 0.697~
(S = 2.11)$.  On the other hand, the solutions in the $\xi^{LR} = \xi^{LL}$
case give $S_{\phi K_S} = -0.83$ for solution $(\mbox{A}_{LR})$ of
Eq.~(\ref{eqn:ARH}) and $S_{\phi K_S} = -0.65$ for solution $(\mbox{B}_{LR})$
of Eq.~(\ref{eqn:BRH}).  Both solutions (A) and (B) satisfy the averaged
experimental $1 \sigma$ range.  As mentioned above, the reason we are able to
find solutions to account for both the $\pi K$ and $S_{\phi K_S}$ anomalies
lies in the fact that the contributions from the $O_{7,8}$ and $O_{9,10}$
operators interfere differently.

\smallskip
{\it Acknowledgments:} C.-W.~C. would like to thank the hospitality of the high
energy theory group at University of Pennsylvania during his visit.  This
research was supported in part by the United States Department of Energy under
Grants No. EY-76-02-3071 and No. DE-FG02-95ER40896, and in part by the
Wisconsin Alumni Research Foundation.



\begin{thebibliography}{99}

\bibitem{Aubert:2003hf}
BABAR Collaboration, B.~Aubert {\it et al.},
Phys.\ Rev.\ Lett.\  91, 241801 (2003).
[hep-ex/0308012].

\bibitem{Chao:2003ue}
Belle Collaboration, Y.~Chao {\it et al.},
BELLE-PREPRINT-2003-26, hep-ex/0311061.

\bibitem{BaBarPlot0054}
BABAR Collaboration, J.~Olsen {\it et al.}, PLOT-0054, \\
{\tt \small https://oraweb.slac.stanford.edu:8080/pls/slac\\
query/BABAR\_DOCUMENTS.DetailedIndex?P\_BP\_ID=3592}.

\bibitem{Abe:2004us}
Belle Collaboration, K.~Abe {\it et al.},
BELLE-PREPRINT-2004-1, hep-ex/0401029.

\bibitem{Buras:0312259}
A.~J.~Buras, R.~Fleischer, S.~Recksiegel and F.~Schwab,
Phys.\ Rev.\ Lett.\  {\bf 92}, 101804 (2004)
[hep-ph/0312259];
CERN-PH-TH-2004-020, hep-ph/0402112.

\bibitem{Chiang:2004nm}
C.~W.~Chiang, M.~Gronau, J.~L.~Rosner and D.~A.~Suprun,
hep-ph/0404073.

\bibitem{BN}
M.~Beneke, G.~Buchalla, M.~Neubert and C.~T.~Sachrajda,
Nucl.\ Phys.\ B 606, 245 (2001);
[hep-ph/0104110];
M.~Beneke and M.~Neubert,
Nucl.\ Phys.\ B 651, 225 (2003);
[hep-ph/0210085];

\bibitem{Beneke:2003zv}
Nucl.\ Phys.\ B 675, 333 (2003);
[hep-ph/0308039];
M.~Neubert, talk given at Super $B$ Factory Workshop, Hawaii, January 19-22,
2004.

\bibitem{Keum:2003qi}
Y.~Y.~Keum and A.~I.~Sanda,
eConf {\bf C0304052}, WG420 (2003)
[hep-ph/0306004].

\bibitem{Bauer:2004tj}
C.~W.~Bauer, D.~Pirjol, I.~Z.~Rothstein and I.~W.~Stewart,
MIT-CTP-3469, hep-ph/0401188.

\bibitem{Aubert:2004xf}
BABAR Collaboration, B.~Aubert {\it et al.},
BABAR-PUB-04-005, hep-ex/0403001.

\bibitem{Buras:2000gc}
A.~J.~Buras and R.~Fleischer,
Eur.\ Phys.\ J.\ C {\bf 16}, 97 (2000)
[hep-ph/0003323];
A.~J.~Buras, R.~Fleischer, S.~Recksiegel and F.~Schwab,
Eur.\ Phys.\ J.\ C {\bf 32}, 45 (2003)
[hep-ph/0309012].

\bibitem{Yoshikawa:2003hb}
T.~Yoshikawa,
Phys.\ Rev.\ D {\bf 68}, 054023 (2003)
[hep-ph/0306147].

\bibitem{Gronau:2003kj}
M.~Gronau and J.~L.~Rosner,
Phys.\ Lett.\ B 572, 43 (2003).
[hep-ph/0307095].

\bibitem{Buras:epjc.11.93}
A.~J.~Buras and R.~Fleischer,
Eur.\ Phys.\ J.\ C {\bf 11}, 93 (1999)
[hep-ph/9810260].

\bibitem{Barger:plb.580.186}
V.~Barger, C.~W.~Chiang, P.~Langacker and H.~S.~Lee,
Phys.\ Lett.\ B {\bf 580}, 186 (2004)
[hep-ph/0310073];
see also
S.~Fajfer and P.~Singer,
Phys.\ Rev.\ D {\bf 65}, 017301 (2002)
[hep-ph/0110233]
for another analysis of constraints on $Z'$ models from a $|\Delta S| = 2$
process.

\bibitem{Barger:2004qc}
V.~Barger, C.~W.~Chiang, J.~Jiang and P.~Langacker,
arXiv:hep-ph/0405108.

\bibitem{Chaudhuri:npb.456.89}
S.~Chaudhuri, S.~W.~Chung, G.~Hockney and J.~Lykken,
Nucl.\ Phys.\ B {\bf 456}, 89 (1995)
[hep-ph/9501361];
G.~Cleaver, M.~Cvetic, J.~R.~Espinosa, L.~L.~Everett, P.~Langacker and J.~Wang,
Phys.\ Rev.\ D {\bf 59}, 055005 (1999)
[hep-ph/9807479];
M.~Cvetic, G.~Shiu and A.~M.~Uranga,
Phys.\ Rev.\ Lett.\ {\bf 87}, 201801 (2001)
[hep-th/0107143];
Nucl.\ Phys.\ B {\bf 615}, 3 (2001)
[hep-th/0107166];
M.~Cvetic, P.~Langacker and G.~Shiu,
Phys.\ Rev.\ D {\bf 66}, 066004 (2002)
[hep-ph/0205252].

\bibitem{Nardi:prd.48.1240}
E.~Nardi,
Phys.\ Rev.\ D {\bf 48}, 1240 (1993)
[hep-ph/9209223];
J.~Bernabeu, E.~Nardi and D.~Tommasini,
Nucl.\ Phys.\ B {\bf 409}, 69 (1993)
[hep-ph/9306251];
Y.~Nir and D.~J.~Silverman,
Phys.\ Rev.\ D {\bf 42}, 1477 (1990);
V.~D.~Barger, M.~S.~Berger and R.~J.~Phillips,
Phys.\ Rev.\ D {\bf 52}, 1663 (1995)
[hep-ph/9503204];
M.~B.~Popovic and E.~H.~Simmons,
Phys.\ Rev.\ D {\bf 62}, 035002 (2000)
[hep-ph/0001302].

\bibitem{Buchalla:prd.53.5185}
G.~Buchalla, G.~Burdman, C.~T.~Hill and D.~Kominis,
Phys.\ Rev.\ D {\bf 53}, 5185 (1996)
[hep-ph/9510376];
G.~Burdman, K.~D.~Lane and T.~Rador,
Phys.\ Lett.\ B {\bf 514}, 41 (2001)
[hep-ph/0012073];
A.~Martin and K.~Lane,
BUHEP-04-03, hep-ph/0404107.

\bibitem{Grossman:1999av}
See also Y.~Grossman, M.~Neubert and A.~L.~Kagan,
JHEP 9910, 029 (1999)
[hep-ph/9909297] for an early analysis of $Z'$ effects on $\pi K$ decays.

\bibitem{Langacker:prd.62.013006}
P.~Langacker and M.~Plumacher,
Phys.\ Rev.\ D {\bf 62}, 013006 (2000)
[hep-ph/0001204].

\bibitem{Buchalla:1995vs}
G.~Buchalla, A.~J.~Buras and M.~E.~Lautenbacher,
Rev.\ Mod.\ Phys.\  {\bf 68}, 1125 (1996)
[hep-ph/9512380].

\bibitem{Babu:1996vt}
K.~S.~Babu, C.~F.~Kolda and J.~March-Russell,
Phys.\ Rev.\ D 54, 4635 (1996)
[hep-ph/9603212];
{\it ibid.} 57, 6788 (1998)
[hep-ph/9710441];
T.~G.~Rizzo,
Phys.\ Rev.\ D 59, 015020 (1999)
[hep-ph/9806397];
K.~Leroux and D.~London,
Phys.\ Lett.\ B 526, 97 (2002)
[hep-ph/0111246].

\bibitem{Fleischer:epjc.6.451}
R.~Fleischer,
Eur.\ Phys.\ J.\ C {\bf 6}, 451 (1999)
[hep-ph/9802433].

\bibitem{Neubert:1998pt}
M.~Neubert and J.~L.~Rosner,
Phys.\ Lett.\ B {\bf 441}, 403 (1998)
[hep-ph/9808493].

\bibitem{Wolfenstein:prl.51.1983}
L.~Wolfenstein,
Phys.\ Rev.\ Lett.\  {\bf 51}, 1945 (1983).

\bibitem{Battaglia:2003in}
M.~Battaglia {\it et al.},
report for the Workshop on CKM Unitarity Triangle (CERN 2002-2003), Geneva,
Switzerland, 13-16 Feb 2002,
hep-ph/0304132.

\bibitem{Abe:prl.91.261602}
K.~Abe {\it et al.}  [Belle Collaboration],
Phys.\ Rev.\ Lett.\  {\bf 91}, 261602 (2003)
[hep-ex/0308035].

\bibitem{Browder:2003ii}
T.~E.~Browder,
Int.\ J.\ Mod.\ Phys.\ A 19, 965 (2004).
[hep-ex/0312024].

\bibitem{Aubert:2004ii}
B.~Aubert {\it et al.}  [BABAR Collaboration],
arXiv:hep-ex/0403026.

\end{thebibliography}
\end{document}